\documentclass[english]{article}
\usepackage[T1]{fontenc}
\usepackage[latin9]{inputenc}
\usepackage{color}
\usepackage{babel}
\usepackage{textcomp}
\usepackage{amsmath}
\usepackage{amssymb}
\usepackage{graphicx}
\usepackage[unicode=true,pdfusetitle,
 bookmarks=true,bookmarksnumbered=false,bookmarksopen=false,
 breaklinks=false,pdfborder={0 0 0},backref=false,colorlinks=true]
 {hyperref}
\usepackage{breakurl}

\makeatletter

\newcommand{\noun}[1]{\textsc{#1}}

\def\NPB#1#2#3{{\em Nucl. Phys.} {\bf B#1} (19#2) #3}
\def\PLB#1#2#3{{\em Phys. Lett.} {\bf B#1} (19#2) #3}

\def\ZPC#1#2#3{{\em Zeit. f\"ur Physik} {\bf C#1} (19#2) #3}

\makeatother

\begin{document}

\title{Higgs Boson Phenomenology in a Simple Model with Vector Resonances}

\author{Oscar Castillo-Felisola, Cristóbal Corral, Marcela González,\\
 Gastón Moreno, Nicolás A. Neill, Felipe Rojas, \\
Jilberto Zamora and Alfonso R. Zerwekh\textit{\normalsize }%
\thanks{\textit{alfonso.zerwekh@usm.cl}%
}\textit{\normalsize }\\
\textit{\normalsize Departamento de Física and }\\
\textit{\normalsize Centro Científico-Tecnológico de Valparaíso}\\
\textit{\normalsize Universidad Técnica Federico Santa María}\\
\textit{\normalsize Casilla 110-V, Valparaíso, Chile}}
\maketitle
\begin{abstract}
In this paper we consider a simple scenario where the Higgs boson
and two vector resonances are supposed to arise from a new strong
interacting sector. We use the ATLAS measurements of the dijet spectrum
to set limits on the masses of the resonances. Additionally we compute
the Higgs boson decay to two photons and found, when compare to the
Standard Model prediction, a small excess which is compatible with
ATLAS measurements. Finally we make prediction for Higgs-strahlung
processes for the LHC running at 14 TeV.
\end{abstract}

\section{Introduction}

The 125 GeV Higgs boson recently discovered at the LHC \cite{Aad:2012tfa,Chatrchyan:2012ufa}
seems to behave in a very standard way. No important deviation from
the Standard Model (SM) predictions has been found except for Mass
2013a small discrepancy in its decay to two photons which still survives
in ATLAS measurements. As a result, stringent constraints have been
put on an eventual New Physics at the TeV scale. The Minimal Supersymmetric
Standard Model (MSSM), for instance, seems to suffer from serious
tensions trying to accommodate simultaneously such a large mass for
the lighter Higgs boson and the current non-observability of any ``super-partner''
at the TeV scale . Indeed, the current exploration of supersymmetric
alternatives to the Standard Model, in general, seems to need a structure
that goes beyond the minimal possibility\cite{Hardy:2013ywa}. On
the other hand, the LHC has evidently ruled out a complete family
of Dynamical Electroweak Symmetry Breaking (DEWSB) models which predicted
a Higgsless low energy spectrum. Nevertheless, there are some classes
of DEWSB models which predict a light composite Higgs boson and still
offer a viable explanation for the stability of the Electroweak scale.
For instance, Walking Technicolor (WTC)\cite{Foadi:2007ue,Ryttov:2008xe,Sannino:2009za}
and models where the Higgs is a pseudo-Goldstone boson are notable
examples of such kind of models with a light scalar in the spectrum\cite{little,compoold,compo}.

In this paper, we consider a scenario where a new strong interacting
sector originates the Fermi scale and manifests itself by means of
a composite scalar boson (which we identify with the 125 GeV Higgs
boson) and vector resonances. In order to make this scenario concrete,
we use a simple phenomenological model proposed by one of us some
year ago\cite{Zerwekh:2005wh}. In this model, two kind of vector
resonances are included: a triplet of $SU(2)_{L}$ (which can be thought
as a ``techni-rho'' and then we denote it by $\rho_{\mu}$) and
a singlet of $SU(2)_{L}$ (which can be thought as a ``techni-omega''
and then we denote it by $\omega_{\mu}$). It is assumed that these
new states interact with the particles of the SM by their mixing with
the standard gauge bosons $W_{\mu}^{a}$ and $B_{\mu}$ in analogy
to the well known Vector Meson Dominance (VMD) mechanism in Hadron
Physics. Originally, this setup was used to point out that in a Composite
Higgs framework an enhancement of the associate production of a Higgs
and a gauge boson could be expected. This result was later confirmed
by other authors in the context of Minimal Walking Technicolor (MWTC)\cite{Belyaev:2008yj}.
It has also been shown that MWTC also predicts a deviation of the
Higgs to two photons decay ratio, compared with the SM. 

In this work, we use the simple framework described above and data
from the ATLAS Collaboration to set limits on the masses of the vector
resonances. Additionally, we make some predictions for the next run
of the LHC. Finally, based on our results, we comment on some characteristics
we think must be respected by DEWSB scenarios.

The paper is organized in the following way. In section \ref{sec:The-Model}
we describe the theoretical construction. Section \ref{sec:Limits-from-the}
is devoted to set limits on the mass of the resonances based on dijet
measurements. In section \ref{sec:hgammagamma} we evaluate the consistency
of the model with the current measurement of $\Gamma(h\rightarrow\gamma\gamma)$.
In section \ref{sec:Higgsstrahlung} we compute Higg-strahlung processes
for the LHC at $\sqrt{s}=14$ TeV. Finally, in section \ref{sec:Conclusion}
we comment our results and state our conclusions.

\section{The Model \label{sec:The-Model}}

The model studied in this work was already proposed elsewhere by one
of us\cite{Zerwekh:2005wh}. Nevertheless, in regard of the completeness,
we dedicate this section to a presentation of the main characteristics
of our theoretical construction.

\subsection{Gauge Sector}

As explained above, our model contains two new vector fields which
are supposed to be the manifestation of an underlying strong sector.
We assume that these resonances interact with the standard sector
through their mixing with the electroweak gauge bosons, following
the VMD idea. In a more modern but equivalent language, the model
can be formulated based on the ``moose'' diagram shown in figure
\ref{fig:Moose-diagram}. 

\begin{center}
\begin{figure}
\begin{centering}
\includegraphics[angle=-90,scale=0.3]{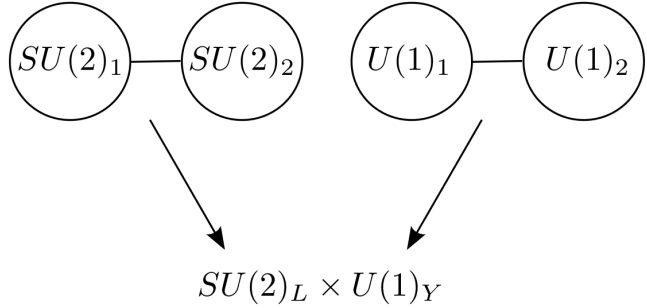}
\par\end{centering}

\caption{Gauge structure of the model. Our construction may be thought as coming
from a moose diagram like the one shown here. The initial gauge symmetry
$SU(2)_{1}\times SU(2)_{2}\times U(1)_{1}\times U(1)_{2}$ is broken
down (by a non-linear sigma sector, for example) to $SU(2)_{L}\times U(1)_{Y}$
and our effective Lagrangian (\ref{eq:Laggauge}) is obtained.\label{fig:Moose-diagram}}
\end{figure}

\par\end{center}

The resulting low energy Lagrangian (assuming that the Abelian and
non-Abelian parts break at the same scale) can be written as:

\begin{eqnarray}
{\cal L} & = & -\frac{1}{2}\mathrm{Tr\left\{ W_{\mu\nu}W^{\mu\nu}\right\} }-\frac{1}{2}\mathrm{Tr}\left\{ \tilde{\rho}_{\mu\nu}\tilde{\rho}^{\mu\nu}\right\} \nonumber \\
 &  & -\frac{1}{4}B_{\mu\nu}B^{\mu\nu}-\frac{1}{4}\tilde{\omega}_{\mu\nu}\tilde{\omega}^{\mu\nu}\nonumber \\
 &  & +M^{2}\mathrm{Tr}\left\{ \left(\frac{g}{g_{2}}W_{\mu}-\tilde{\rho}_{\mu}\right)^{2}\right\} +\frac{M^{2}}{2}\left(\frac{g^{\prime}}{g{}_{2}}B_{\mu}-\tilde{\omega}_{\mu}\right)^{2}.\label{eq:Laggauge}
\end{eqnarray}
where $W_{\mu}$ and $B_{\mu}$ are the fields associated to the sites
labeled by ``1'' while $\tilde{\rho}_{\mu}$ and $\tilde{\omega}_{\mu}$
are associated to the sites labeled by ``2'' and assumed to be strongly
coupled. As usual, $W_{\mu\nu}$ , $\tilde{\rho}_{\mu\nu}$, $B_{\mu\nu}$
and $\tilde{\omega}_{\mu\nu}$ are (in term of components)

\begin{eqnarray}
W_{\mu\nu}^{a} & = & \partial_{\mu}W_{\nu}^{a}-\partial_{\nu}W_{\mu}^{a}+g\epsilon^{abc}W_{\mu}^{b}W_{\nu}^{c},\label{eq:def1}\\
\tilde{\rho}_{\mu\nu}^{a} & = & \partial_{\mu}\tilde{\rho}_{\nu}^{a}-\partial_{\nu}\tilde{\rho}_{\mu}^{a}+g_{2}\epsilon^{abc}\tilde{\rho}_{\mu}^{b}\tilde{\rho}_{\nu}^{c},\\
B_{\mu\nu} & = & \partial_{\mu}B_{\nu}-\partial_{\nu}B_{\mu},\\
\tilde{\omega}_{\mu\nu} & = & \partial_{\mu}\tilde{\omega}_{\nu}-\partial_{\nu}\tilde{\omega}_{\mu}.
\end{eqnarray}

By construction, Lagrangian (\ref{eq:Laggauge}) is invariant under
$SU(2)_{\mbox{L}}\times U(1)_{\mbox{Y}}$. The symmetry breaking to
$U(1)_{\mbox{em}}$ will be described by means of the vacuum expectation
value of a scalar field, as in the Standard Model. In other words,
we will use an effective gauged linear sigma model as a phenomenological
description of the electroweak symmetry breaking.

\subsection{Fermion Sector}

As usual, fermions are coupled to the gauge fields through covariant
derivatives:

\begin{equation}
{\cal L}=\overline{\psi_{L}}i\gamma^{\mu}D_{\mu}\psi_{L}+\overline{\psi_{R}}i\gamma^{\mu}\tilde{D}_{\mu}\psi_{R},\label{eq:fermionlag}
\end{equation}
with 
\begin{eqnarray}
D_{\mu} & = & \partial_{\mu}+i\tau^{a}g(1-x_{1})W_{\mu}^{a}+i\tau^{a}g_{2}x_{1}\tilde{\rho}_{\mu}^{a}\nonumber \\
 &  & +i\frac{Y}{2}g'(1-x_{2})B_{\mu}+i\frac{Y}{2}g'_{2}x_{2}\tilde{\omega}_{\mu}
\end{eqnarray}
and 
\begin{equation}
\tilde{D}_{\mu}=\partial_{\mu}+i\frac{Y}{2}g'(1-x_{3})B_{\mu}+i\frac{Y}{2}g'_{2}x_{3}\tilde{\omega}_{\mu}.
\end{equation}

The parameters $x_{i}$ ($i=1,2,3$) play the role of fermion delocalization
\cite{Chivukula:2005bn} and in our case govern the direct coupling
to the new vector bosons. In the context of the BESS model \cite{dominici,mbess},
which shares with our model a similar structure in the spin-1 sector,
it has been shown that a direct interaction of order $2(g/g_{2})^{2}$
is necessary in order to reconcile the model with the precision electroweak
parameters $S$, $T$ and $U$. Hence, we adopt this result and impose
$x_{i}=2(g/g_{2})^{2}$ for $i=1,2,3$.

\subsection{Higgs Sector and EWSB}

In our effective model, the Higgs sector is assumed to be the same
as in the Standard Model except by the possibility of including a
direct coupling, between the Higgs doublet and the vector resonances,
depending on how the Higgs boson is localized. In this way, the Lagrangian
for the Higgs sector can be written as:

\begin{equation}
{\cal L}=\left(D^{\mu}\Phi\right)^{\dagger}\left(D_{\mu}\Phi\right)-V\left(\Phi\right),\label{eq:Higgs}
\end{equation}
where, as usual

\begin{equation}
V\left(\Phi\right)=-\mu^{2}\Phi^{\dagger}\Phi+\lambda\left(\Phi^{\dagger}\Phi\right)^{2},\label{eq:potential}
\end{equation}
and

\begin{eqnarray}
D_{\mu} & = & \partial_{\mu}+i\tau^{a}g(1-f_{1})W_{\mu}^{a}+i\tau^{a}g_{2}f_{1}\tilde{\rho}_{\mu}^{a}\nonumber \\
 &  & +i\frac{Y}{2}g'(1-f_{2})B_{\mu}+i\frac{Y}{2}g'_{2}f_{2}\tilde{\omega}_{\mu}.\label{eq:HiggsCovDeriv}
\end{eqnarray}

For simplicity, we chose to completely localize the Higgs fields in
the weakly coupled sites imposing $f_{i}=0$ for $i=1,2$. The consequences
of these choice will be discussed later.

Once the electroweak symmetry is broken, two non-diagonal mass matrices
are generated: one for the neutral vector bosons and another for the
charged one 

\begin{equation}
{\cal M_{\mbox{neutral}}}=\frac{v^{2}}{4}\left[\begin{array}{cccc}
(1+\alpha^{2})g^{2} & -\alpha^{2}gg_{2} & -gg^{\prime} & 0\\
-\alpha^{2}gg_{2} & \alpha^{2}g_{2}^{2} & 0 & 0\\
-gg^{\prime} & 0 & (1+\alpha^{2})g{}^{\prime2} & -\alpha^{2}g^{\prime}g_{2}\\
0 & 0 & -\alpha^{2}g^{\prime}g_{2} & \alpha^{2}g{}^{\prime2}
\end{array}\right],\label{eq:neutralmassmatrix}
\end{equation}

\begin{equation}
{\cal M_{\mbox{charged}}}=\frac{v^{2}}{4}\left[\begin{array}{cc}
(1+\alpha^{2})g^{2} & -\alpha^{2}gg_{2}\\
-\alpha^{2}gg_{2} & \alpha^{2}g_{2}^{2}
\end{array}\right],\label{eq:chargedmassmatrix}
\end{equation}

where

\begin{equation}
\alpha^{2}=\frac{4M^{2}}{v^{2}g_{2}^{2}}.\label{eq:alpha}
\end{equation}

Notice that $\alpha v$ may be thought as the scale where the mass
of the rho and omega resonances is generated. It is natural to identify
this scale with the natural scale of the strong sector which in our
case is $\Lambda=4\pi v$. Consequently, we assume that the natural
value of $\alpha$ is $\alpha=4\pi$. Notice that $\Lambda$ is also
a natural upper limit for the value of the mass of the resonances.
Additionally, we have assume that the new vector resonance are degenerated
in mass. 

The mass matrices can be diagonalized and the eigenvectors, in the
limit $g/g_{2}\ll1$ and keeping terms up to order $g/g_{2}$, can
be written as:

\begin{eqnarray}
A & = & \frac{g'}{\sqrt{g^{2}+g'^{2}}}W^{3}+\frac{g}{\sqrt{g^{2}+g'^{2}}}B+\frac{gg'}{g_{2}\sqrt{g^{2}+g'^{2}}}\tilde{\rho}^{3}+\frac{gg'}{g_{2}\sqrt{g^{2}+g'^{2}}}\tilde{\omega},\nonumber \\
Z & = & \frac{g}{\sqrt{g^{2}+g'^{2}}}W^{3}-\frac{g'}{\sqrt{g^{2}+g'^{2}}}B+\frac{g^{2}}{g_{2}\sqrt{g^{2}+g'^{2}}}\tilde{\rho}^{3}-\frac{g'^{2}}{g_{2}\sqrt{g^{2}+g'^{2}}}\tilde{\omega},\nonumber \\
\rho^{0} & = & -\frac{g}{g_{2}}W^{3}+\tilde{\rho}^{3},\\
\omega & = & -\frac{g'}{g_{2}}B+\tilde{\omega},\nonumber \\
W^{\pm} & = & \tilde{W}^{\pm}+\frac{g}{g_{2}}\tilde{\rho}^{\pm},\nonumber \\
\rho^{\pm} & = & \tilde{\rho}^{\pm}-\frac{g}{g_{2}}\tilde{W}^{\pm},\nonumber 
\end{eqnarray}
where 
\begin{equation}
\tilde{W}^{\pm}=\frac{1}{\sqrt{2}}\left(W^{1}\mp iW^{2}\right)
\end{equation}
and 
\begin{equation}
\tilde{\rho}^{\pm}=\frac{1}{\sqrt{2}}\left(\tilde{\rho}^{1}\mp i\tilde{\rho}^{2}\right).
\end{equation}

We used LanHEP \cite{Semenov:2010qt} to implement this model into
the CalcHEP package\cite{Belyaev:2012qa}.

\section{Limits from the Dijet Spectrum\label{sec:Limits-from-the}}

\begin{figure}
\begin{centering}
\includegraphics[scale=0.5]{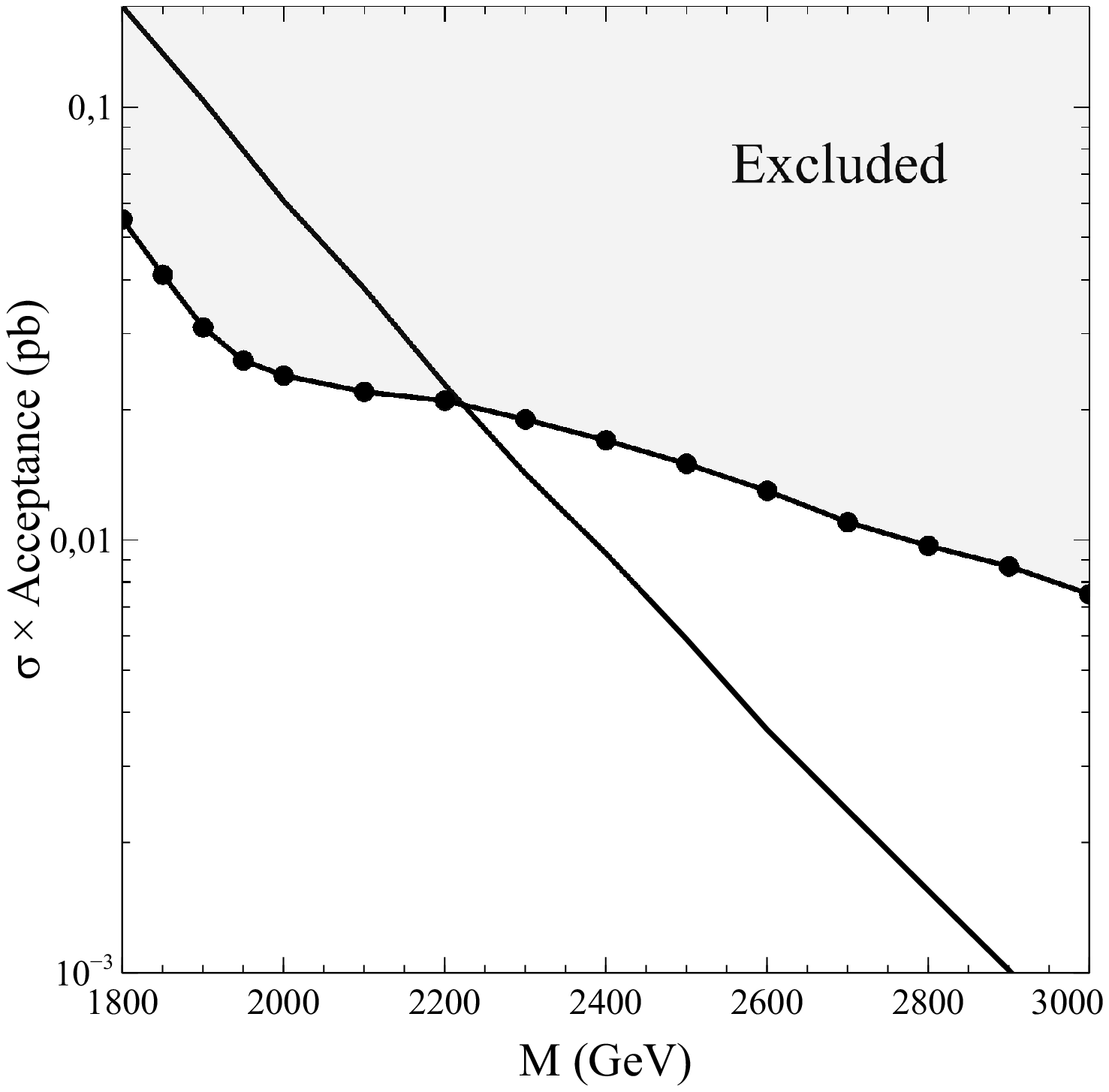}
\par\end{centering}

\caption{Resonant cross sections (continuous line) compared to ATLAS upper
limits to narrow resonances decaying into dijet (dots with continuous
line), as a function of the resonance mass. The colored region is
excluded. \label{fig:dijets}}
\end{figure}

As a first step, we proceed to obtain limits for the mass of the resonances
using the dijet spectrum measured by ATLAS\cite{ATLAS:2012qjz}. For
this purpose, we generated events for the $pp\rightarrow jj$ process
at $\sqrt{s}=8$ TeV using CalcHEP, considering only the contribution
of the new vector resonances. In order to compare with ATLAS data
we imposed the following kinematic cuts:

\begin{eqnarray}
\frac{1}{2}\left|y_{1}-y_{2}\right| & < & 0.6,\\
\left|y_{1,2}\right| & < & 2.8,\\
M_{jj} & > & 1000\:\mathrm{GeV,}\\
p_{T1,2} & > & 150\:\mathrm{GeV,}
\end{eqnarray}
where $y$ refers to the jet rapidity, $p_{T}$ is its transverse
momentum and $M_{jj}$ is the invariant mass of the two jets. The
main contribution to this process comes from the production of the
rho and omega in the s-channel through quark--anti-quark annihilation.
Unfortunately, due to PDF effects, it is unlikely to produce anti-quarks
with large momentum at the LHC and a considerable fraction of the
events are produced in the low $M_{jj}$ region. Of course, such events
are hidden under the background. In consequence we select only the
resonant part of our events in order to compare it with the experimental
upper limits. For doing that, and after imposing the cuts, we fit
the events around the resonant peak with a Breit-Wigner function and
a second degree polynomial, describing the non-resonant ``background'',
and define our resonant events as the integral of the Breit-Wigner
function. Now we are prepared to compare with experimental data. Our
results are shown in figure \ref{fig:dijets}. There, the ATLAS upper
limits for invisible resonances are compared to our predicted cross
section for dijet production considering only the contribution of
the new vector fields and the appropriate kinematic cuts. Everything
above the experimental curve is excluded. In our case, that means
that our model is consistent with the ATLAS dijet measurements provided
that the mass of the vector resonances satisfies the constrain $M\gtrsim2200$
GeV. 

\section{$h\rightarrow\gamma\gamma$\label{sec:hgammagamma}}

\begin{figure}
\begin{centering}
\includegraphics[scale=0.5]{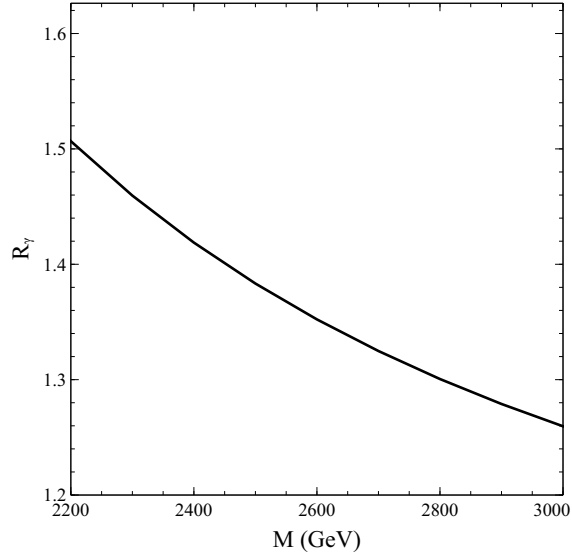}
\par\end{centering}

\caption{Predicted values of $R_{\gamma}=\Gamma(h\rightarrow\gamma\gamma)_{Model}/\Gamma(h\rightarrow\gamma\gamma)_{SM}$
\label{fig:h-to-A-A}}
\end{figure}

A relevant observable in Higgs phenomenology is its decay rate into
two photons. This is the only measured quantity that presents a sensible
deviation from the SM, at least in results reported by ATLAS. Since
it is a loop effect, we expect that this observable is sensible to
the existence of New Physics. With this motivation in mind, we computed
the $\Gamma(h\rightarrow\gamma\gamma)$ in our model at one loop level.
Non-standard contributions arise through the presence of the $\rho_{\mu}^{\pm}$
fields. The new diagrams are formally the same that those generated
by the $W_{\mu}^{\pm}$ bosons except, obviously, by the different
values of the coupling constants (which in case of the $h\rho^{+}\rho^{-}$
is suppressed by a factor $(g/g_{2})^{2}$ with respect to $hW^{+}W^{-}$)
and the vector boson masses. The result is shown in figure \ref{fig:h-to-A-A}
where we have plotted the ratio

\begin{equation}
R_{\gamma}=\frac{\Gamma(h\rightarrow\gamma\gamma)_{Model}}{\Gamma(h\rightarrow\gamma\gamma)_{SM}}
\end{equation}
as a function of the rho mass. We see that in the allowed range of
masses, $R_{\gamma}$ varies from 1.25 to 1.50. This result has to
be compared with the experimental values\cite{AtlashAA,CMShAA}:

\begin{equation}
R_{\gamma}=\begin{cases}
1.6\pm0.3 & \mathrm{ATLAS}\\
0.77\pm0.27 & \mathrm{CMS}.
\end{cases}
\end{equation}

Notice that our results agree with the experimental values at $1\sigma$
level in the case of ATLAS and at $2.5\sigma$ level in the case of
CMS 

\section{Higgs-strahlung\label{sec:Higgsstrahlung}}

\begin{figure}
\begin{centering}
\includegraphics[scale=0.5]{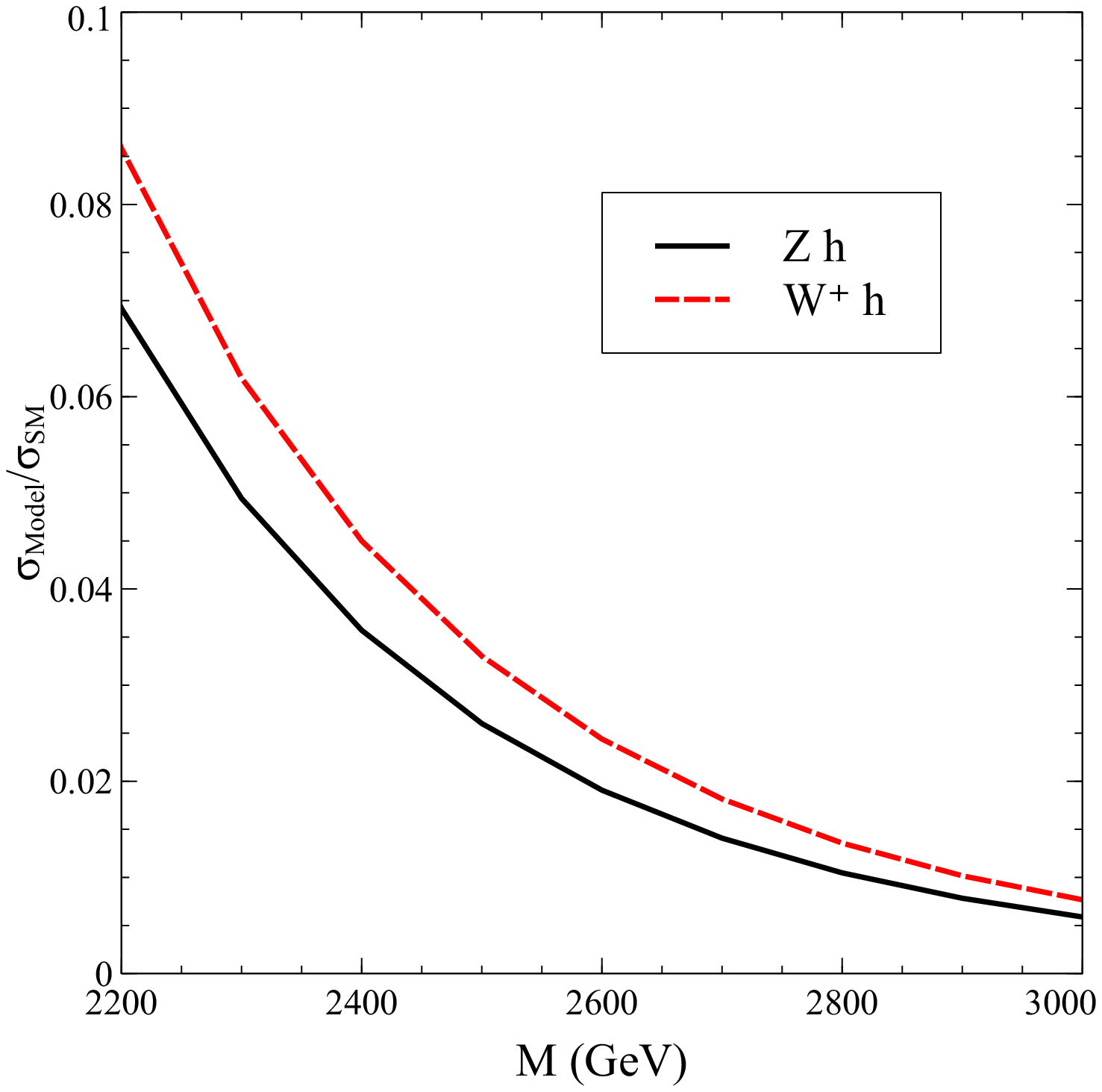}
\par\end{centering}

\caption{Predicted cross sections for the processes $pp\rightarrow Zh$ (continuous
line) and $pp\rightarrow W^{+}h$ (dashed line) at $\sqrt{s}=14$
TeV due only to the contributions of the new spin-1 states compared
to the SM prediction\label{fig:hv}}
\end{figure}

As explained above, the model studied in this paper was proposed by
one of us some years ago. It was shown that for resonance masses that
were reasonable at that time, the model presented a significant enhancement
of associated production of a Higgs boson and a gauge boson. This
result was confirmed later in the framework of the MWTC. Given this
context, it is interesting to evaluate whether this channel remains
robust after considering the limits imposed by current experimental
data. Consequently, we used CalcHEP to compute the cross sections
for the processes $pp\rightarrow Zh$ and $pp\rightarrow W^{+}h$
at $\sqrt{s}=14$ TeV due only to the contributions of the new spin-1
states and we compare it to the SM prediction. The result is shown
in figure \ref{fig:hv}.

Unfortunately, our results show a very small enhancement in the cross
section: less than $10\%$ in the available mass range, which is unlikely
to be testable at the LHC.

\section{Summary and Conclusions\label{sec:Conclusion}}

In this work we have reconsidered a simple model which contains two
new spin-1 states, supposed to be composite together with the Higgs
boson, pointing out to a strong dynamical origin of the Electroweak
scale. We used dijet data from ATLAS to set limits on the mass of
the new states. We found that current data allows masses in the range
$2.2\:\mathrm{TeV}\lesssim M\lesssim4\pi v\approx3.1$ TeV . 

Then, we computed $\Gamma(h\rightarrow\gamma\gamma)$ and compared
with the SM prediction. We found that our result is compatible with
recent measurements at the LHC. In this sense we can say that this
simple model has passed the experimental test. This success is related
to the fact that in our model the Higgs boson is weakly coupled to
the new spin-1 states. This is a consequence of choosing $f_{1}=f_{2}=0$
in equation (\ref{eq:HiggsCovDeriv}). The original motivation for
this choice was to simplify the model and make sure we could faithfully
reproduce the phenomenology of the $W^{\pm}$ and $Z$ bosons. Nevertheless,
we see the agreement with the measured value of $R_{\gamma}$ as an
\emph{a posteriori} justification.\emph{ }

Unfortunately, on the other hand, the proposed enhancement in the
Higgs-strahlung channels has not survived as an important signal of
the model at the LHC.

\section*{Ackowledgements}

ARZ has received financial support from Fondecyt grant nº 1120346.
O.C-F. was supported by Fondecyt grant No. 11000287 and Basal Project
FB0821. CC, GM, FR and JZ received support from Conicyt National Ph.D.
Fellowship Program.

\end{document}